\documentclass{issfd2026}

\usepackage{fancyhdr}

\usepackage{etoolbox}

\makeatletter
\patchcmd{\thebibliography}{\section{\refname}}{}{}{}
\patchcmd{\thebibliography}{\section*{\refname}}{}{}{}
\makeatother

\makeatletter
\let\ISSFD@oldthebibliography\thebibliography
\let\ISSFD@oldendthebibliography\endthebibliography

\renewenvironment{thebibliography}[1]
  {%
    \ISSFD@oldthebibliography{#1}%
    \setlength{\itemsep}{-1pt}%
    \setlength{\parsep}{0pt}%
    \setlength{\parskip}{0pt}%
    \setlength{\partopsep}{0pt}%
    \setlength{\topsep}{0pt}%
  }
  {%
    \ISSFD@oldendthebibliography
  }
\makeatother

\usepackage{bm}
\usepackage{amsmath}
\usepackage{etoolbox}
\usepackage{booktabs}
\usepackage{tabularx}

\ISSFDtitle{Efficient Nonlinear Uncertainty Quantification for Spaceflight Leveraging Nonlinear Expansions}
\ISSFDauthors{Ethan R. Burnett\textsuperscript{(1)}, Spencer Boone\textsuperscript{(2)}} 
\ISSFDaffiliations{%
\textsuperscript{(1)}University of Texas at Austin\\
Austin, TX, United States\\
Email: ethan.burnett@utexas.edu\\[12pt]
\textsuperscript{(2)} ISAE-SUPAERO \\
Toulouse, France\\
Email: spencer.boone@isae-supaero.fr}
\ISSFDabstract{This paper provides a comparative study of modern uncertainty quantification (UQ) methods. To greatly enhance real-time performance, both differential algebra (DA) and a directional differential algebra (DDA) approach are employed. This can enable fast UQ in the case of non-Gaussian statistics. Higher-order moments, namely skew and kurtosis, can be computed quickly by several means. This motivates their implementation in an analytic approximation of the confidence bounds for the so-called “banana-shaped" non-Gaussian distributions encountered often in nonlinear astrodynamics problems. This method improves greatly on a linear covariance approach, with only $\sim$5$\times$ its runtime in numerical tests, even before DA methods are employed. Test problems in this work include a restricted three-body cislunar example and an Earth-return aerocapture example.}

\begin{document}
\maketitle

\section{Introduction}
Reliable and efficient uncertainty quantification (UQ) is a persistent challenge in modern astrodynamics, especially for nonlinear and weakly stable regimes. Modern astrodynamics encompasses cislunar flight, close-proximity operations around small-bodies, entry, descent, and landing (EDL) and related aerobraking and aerocapture problems. In these environments, initial small Gaussian distributions can evolve into non-Gaussian shapes due to perturbations, nonlinear dynamics, and resonances \cite{scheeres2016orbital,wolf_multi-fidelity_2022}. Accurate geometric description of these distributions may depend not only on mean and covariance, but also on higher-order moments such as skewness and kurtosis. Standard Monte Carlo (MC) methods present high-fidelity benchmarks, but are often too computationally expensive for rapid iteration or onboard deployment. Conversely, first-order linear covariance propagation is computationally efficient but can substantially misrepresent nonlinear uncertainty transport \cite{michelotti_uncertainty_2024}.

Sigma-point methods and Polynomial Chaos Expansion (PCE) provide an intermediate level of fidelity and computational cost \cite{luo2017review}. The Unscented Transform (UT) captures nonlinear mean and covariance propagation, while Conjugate Unscented Transform \cite{CUT_ACC}  enables direct higher-moment extraction. However, standard sample-based methods still require repeated dynamics evaluations for each realization, which can become very computationally expensive with repetitive UQ queries, which occur in the context of trade studies or in the inner loop of stochastic optimization schemes.

In this context, Differential Algebra (DA) techniques \cite{valli_nonlinear_2013} offer a powerful way to compute high-order Taylor expansions of the flow map, enabling repeated UQ evaluations without re-integrating the full dynamics. Furthermore one need not expand in all problem variables, and careful selection of the expansion variables can drastically reduce the DA overhead cost. This approach could be termed ``directional differential algebra", in reference to the mathematically analogous ``directional state transition tensors" defined in \cite{Boone_STTorbits}. Such an approach enables rapid computation of low-order statistics such as the mean and covariance, but its directional structure inherently leads to degraded recovery of higher-order mixed central moments. 

This work compares different methods for non-Gaussian uncertainty approximation and, in particular, for high-order uncertainty quantification in nonlinear astrodynamics. The study highlights trades in accuracy and computational efficiency. The information contained in the estimated higher moments is then used to reconstruct non-elliptical confidence geometry through an analytical banana approximation, which captures moderate non-Gaussian deformation effects \cite{BurnettBooneCDC26}.

\section{Theory and Background}
\subsection{Review of UQ Methods}

We briefly review several uncertainty quantification (UQ) strategies used for nonlinear
state propagation. Let $\bm{X}_0\in\mathbb{R}^N$ denote the random initial state, and let
$\bm{X}_r$ denote the reference state about which a local approximation is constructed.
We define the initial deviation as
\begin{equation}
\label{eq:initial_deviation_reference}
\delta\bm{X}_0 = \bm{X}_0-\bm{X}_r .
\end{equation}
The propagated final state is written as
\begin{equation}
\label{eq:flow_map_def}
\bm{X}_f = \bm{\varphi}_{t_f,t_0}(\bm{X}_0)  
\end{equation}
where $\bm{\varphi}_{t_f,t_0}$ denotes the nonlinear flow map from $t_0$ to $t_f$.
More generally, for a nonlinear output map $\bm{g}:\mathbb{R}^N\rightarrow\mathbb{R}^m$,
we write
\begin{equation}
\label{eq:generic_output_map_def}
\bm{Y} = \bm{g}(\bm{X}).
\end{equation}
Most classical
spaceflight UQ methods focus on the propagated mean and covariance,
\begin{equation}
\label{eq:mean_cov_def_review}
\begin{split}
\bm{\mu}_{f}
&=
\mathbb{E}[\bm{X}_f],\\
P_{f}
&=
\mathbb{E}\!\left[
(\bm{X}_f-\bm{\mu}_{f})
(\bm{X}_f-\bm{\mu}_{f})^\top
\right],
\end{split}
\end{equation}
though non-Gaussian uncertainty characterization also motivates recovery of higher-order
central moments.

\subsubsection{Monte Carlo}

Monte Carlo (MC) propagation approximates the final distribution by directly sampling
the initial distribution and propagating each sample through the nonlinear dynamics. Given
samples $\bm{X}_{0,i}$, $i=1,\ldots,N_s$, the propagated samples are
\begin{equation}
\label{eq:mc_sample_propagation_review}
\bm{X}_{f,i}=\bm{\varphi}_{t_f,t_0}(\bm{X}_{0,i}).
\end{equation}
The sample mean and covariance are then computed as
\begin{equation}
\label{eq:mc_mean_cov_review}
\begin{split}
\hat{\bm{\mu}}_f
&=
\frac{1}{N_s}
\sum_{i=1}^{N_s}\bm{X}_{f,i},\\
\hat{P}_f
&=
\frac{1}{N_s-1}
\sum_{i=1}^{N_s}
(\bm{X}_{f,i}-\hat{\bm{\mu}}_f)
(\bm{X}_{f,i}-\hat{\bm{\mu}}_f)^\top .
\end{split}
\end{equation}
Higher-order central moments are estimated similarly by replacing the outer product in Eq.~\eqref{eq:mc_mean_cov_review} with higher-order tensor products of the centered samples. MC is broadly applicable and often serves as a high-fidelity reference, with sample statistics asymptotically converging to the true statistics, but its accuracy requires a rather high sample count -- worse for off-diagonal covariance components or mixed central moment terms. It can therefore be computationally expensive when each sample requires a full nonlinear propagation \cite{luo2017review}.

\subsubsection{Linear Covariance}

Linear covariance propagation approximates the nonlinear flow locally by its first-order
state transition matrix (STM),
\begin{equation}
\label{eq:stm_flow_derivative_review}
\Phi(t_f,t_0)
=
\frac{\partial \bm{\varphi}_{t_f,t_0}}{\partial \bm{X}}
\bigg|_{\bm{X}_r}.
\end{equation}
For deviations from the reference trajectory,
\begin{equation}
\label{eq:lincov_perturbation_review}
\delta\bm{X}(t_f)
\approx
\Phi(t_f,t_0)\delta\bm{X}(t_0),
\end{equation}
and the covariance is propagated as
\begin{equation}
\label{eq:lincov_cov_review}
P_{f}
=
\Phi(t_f,t_0)
P_{0}
\Phi^{\top}(t_f,t_0).
\end{equation}
This approach is computationally extremely fast and is often adequate when the uncertainty is small and the flow is sufficiently linear in the region and timescale of interest. However, because the distribution is represented only by its first two moments under a linearized map, LinCov is not suitable as a method when nonlinearity and/or non-Gaussianity become important \cite{michelotti_uncertainty_2024}.

\subsubsection{Unscented Transformation}

The unscented transformation (UT) replaces random sampling with a deterministic set of
sigma points designed to capture the input mean and covariance. For a generic nonlinear
map $\bm{g}$, sigma points $\bm{\mathcal{X}}_i$, and weights $w_i$, the quadrature form is
\begin{equation}
\label{eq:ut_quadrature_generic_review}
\mathbb{E}[\bm{g}(\bm{X})]
\approx
\sum_i w_i \bm{g}(\bm{\mathcal{X}}_i).
\end{equation}
For state propagation, one chooses $\bm{g}=\bm{\varphi}_{t_f,t_0}$, so that
\begin{equation}
\label{eq:ut_sigma_propagation_review}
\bm{X}_{f,i}
=
\bm{\varphi}_{t_f,t_0}(\bm{\mathcal{X}}_i).
\end{equation}
The propagated mean and covariance are then approximated as
\begin{equation}
\label{eq:ut_mean_cov_review}
\begin{split}
\bm{\mu}_{f}
&\approx
\sum_i w_i^{(m)}\bm{X}_{f,i},\\
P_{f}
&\approx
\sum_i w_i^{(c)}
(\bm{X}_{f,i}-\bm{\mu}_{f})
(\bm{X}_{f,i}-\bm{\mu}_{f})^\top .
\end{split}
\end{equation}
The UT is much less expensive than MC, requiring only $2N+1$ propagated points in its
standard form, and typically captures important nonlinear corrections to the mean and
covariance. It does not, however, directly target recovery of all higher-order mixed central
moments \cite{unscented}.

\subsubsection{Conjugate Unscented Transformation}

The conjugate unscented transformation (CUT) extends the sigma-point idea by adding conjugate directions to improve recovery of higher-order moments. In the fourth-order
variant, ``CUT4", the sigma set includes the usual axis-aligned points together with off-axis
conjugate points, leading to
\begin{equation}
\label{eq:cut_num_points_review}
N_{\mathrm{CUT4}} = 1+2N+2^N
\end{equation}
points before optional removal of the zero-weight center point. Once the points are
propagated through the flow, the third- and fourth-order central moments may be approximated as
\begin{equation}
\label{eq:cut_high_order_review}
\begin{split}
M^{(3)}_{abc}
&\approx
\sum_i w_i
(X_{i,a}-\mu_{a})
(X_{i,b}-\mu_{b})
(X_{i,c}-\mu_{c}),\\
M^{(4)}_{abcd}
&\approx
\sum_i w_i
(X_{i,a}-\mu_{a})
(X_{i,b}-\mu_{b})\\
&\hspace{2.2em}\times
(X_{i,c}-\mu_{c})
(X_{i,d}-\mu_{d}),
\end{split}
\end{equation}
where the ``$f$" subscripts are dropped for compactness. CUT therefore provides a means for higher-moment recovery at substantially
lower cost than large-sample MC. It also has the strong advantage of being a deterministic scheme for moment estimation. For the full details of implementation, consult Reference \cite{CUT_ACC}.

\subsubsection{Polynomial Chaos Expansion}
Polynomial chaos expansion (PCE) approximates a random output as a finite expansion in
polynomials of simpler random variables. Let
$\bm{\xi}=(\xi_1,\ldots,\xi_{n_\xi})^\top$ denote a vector of standardized independent
random variables, and let
$\bm{\alpha}=(\alpha_1,\ldots,\alpha_{n_\xi})$ denote a
multi-index, where each $\alpha_k$ is a nonnegative integer.. The associated multivariate polynomial basis function is
\begin{equation}
\label{eq:pce_basis_definition_review}
\Psi_{\bm{\alpha}}(\bm{\xi})
=
\prod_{k=1}^{n_\xi}
\psi_{\alpha_k}(\xi_k),
\end{equation}
where $\psi_{\alpha_k}$ is a univariate polynomial of degree $\alpha_k$. The family
$\{\Psi_{\bm{\alpha}}\}$ is chosen to be orthogonal under expectation. That is,
\begin{equation}
\label{eq:pce_orthogonality_review}
\begin{split}
\mathbb{E}\!\left[
\Psi_{\bm{\alpha}}(\bm{\xi})
\Psi_{\bm{\beta}}(\bm{\xi})
\right]
&=
h_{\bm{\alpha}}\delta_{\bm{\alpha}\bm{\beta}},\\
h_{\bm{\alpha}}
&=
\mathbb{E}\!\left[
\Psi_{\bm{\alpha}}^2(\bm{\xi})
\right],
\end{split}
\end{equation}
where $h_{\bm{\alpha}}$ is the squared norm of the basis function and
$\delta_{\bm{\alpha}\bm{\beta}}$ is the Kronecker delta. If an orthonormal basis is used,
then $h_{\bm{\alpha}}=1$. For Gaussian random variables, the corresponding polynomial
family is the Hermite polynomial basis.

For a generic nonlinear map $\bm{g}$ and random input $\bm{X}(\bm{\xi})$, PCE
approximates the output $\bm{Y}=\bm{g}(\bm{X}(\bm{\xi}))$ as
\begin{equation}
\label{eq:pce_generic_expansion_review}
\begin{split}
\bm{Y}
&=
\bm{g}(\bm{X}(\bm{\xi}))\\
&\approx
\sum_{\bm{\alpha}\in\mathcal{A}}
\bm{c}_{\bm{\alpha}}
\Psi_{\bm{\alpha}}(\bm{\xi}),
\end{split}
\end{equation}
where $\mathcal{A}$ is the retained set of multi-indices and
$\bm{c}_{\bm{\alpha}}$ are vector-valued coefficients. 

For state propagation, the same idea is applied to the flow map. Writing the random
initial state as
$\bm{X}_0(\bm{\xi})=\bm{X}_r+\delta\bm{X}_0(\bm{\xi})$, the propagated state is
approximated by
\begin{equation}
\label{eq:pce_flow_expansion_review}
\begin{split}
\bm{X}_f
&=
\bm{\varphi}_{t_f,t_0}
\big(\bm{X}_r+\delta\bm{X}_0(\bm{\xi})\big)\\
&\approx
\sum_{\bm{\alpha}\in\mathcal{A}}
\bm{c}_{\bm{\alpha}}
\Psi_{\bm{\alpha}}(\bm{\xi}).
\end{split}
\end{equation}
The coefficients $\bm{c}_{\bm{\alpha}}$ may be obtained non-intrusively e.g. via sampling and regression. Once these coefficients are known, the moments follow directly
from orthogonality. In particular, because $\Psi_{\bm{0}}=1$ is the constant basis
function, the propagated mean and covariance are
\begin{equation}
\label{eq:pce_mean_review}
\bm{\mu}_f
\approx
\bm{c}_{\bm{0}}.
\end{equation}
\begin{equation}
\label{eq:pce_covariance_review}
P_f
\approx
\sum_{\bm{\alpha}\in\mathcal{A}\setminus\{\bm{0}\}}
h_{\bm{\alpha}}\,
\bm{c}_{\bm{\alpha}}
\bm{c}_{\bm{\alpha}}^\top .
\end{equation}
Thus, after the PCE coefficients are computed, mean and covariance recovery is algebraic.
Higher-order central moments can also be computed from products of basis functions, though
the bookkeeping becomes more involved. Overall, PCE provides an accurate and intermediate-cost
method between direct MC sampling and LinCov \cite{luo2017review, valli_nonlinear_2013, JonesPCE}.

\subsubsection{Gaussian Mixture Model}

A Gaussian mixture model (GMM) represents a non-Gaussian distribution as a weighted
sum of Gaussian components,
\begin{equation}
\label{eq:gmm_density_review}
\begin{split}
p_{\bm{X}_0}(\bm{x})
&\approx
\sum_{\ell=1}^{N_g}
w_{\ell}\,
\mathcal{N}(\bm{x};\bm{\mu}_{0,\ell},P_{0,\ell}),\\
\sum_{\ell=1}^{N_g}w_{\ell}
&=
1.
\end{split}
\end{equation}
Each Gaussian component can then be propagated using any of the preceding local UQ
methods, with LinCov being the fastest. After propagating the $\ell$th Gaussian component, let $\bm{\mu}_{f,\ell}$ and $P_{f,\ell}$ denote the resulting mean and covariance. Then,
\begin{subequations}
\label{eq:gmm_moment_matching_review}
\begin{align}
\bm{\mu}_{f}
&=
\sum_{\ell=1}^{N_g}
w_{\ell}\bm{\mu}_{f,\ell},\\
\begin{split}
P_{f}
&=
\sum_{\ell=1}^{N_g}
w_{\ell}
\left[
P_{f,\ell}
+
(\bm{\mu}_{f,\ell}-\bm{\mu}_{f})
\right.\\
&\hspace{2.2em}\left.
\times
(\bm{\mu}_{f,\ell}-\bm{\mu}_{f})^\top
\right].
\end{split}
\end{align}
\end{subequations}
This makes GMMs attractive for nonlinear problems, as nonlinear deformation can be approximated by the collective evolution of several local Gaussian components. In this work, the mixture construction is included for completeness, but is not covered in detail. The assigning of weights, the mixture ``splitting" process, and the risk allocation procedure for a chance-constrained approach are all rather algorithmically complex and discussed elsewhere. See Reference \cite{boone_cdc} and references therein. Also, to the familiar reader, the moment reconstruction procedure detailed above might seem rather indirect: In this work, we will use higher moments to predict geometric features of a distribution at a particular confidence level. However, GMM is inherently more ``geometric" than the previously mentioned techniques, being well-suited for direct confidence boundary construction without needing to explicitly recover moments. 

\subsection{Differential Algebra}

Differential algebra (DA) \cite{Berz_ParticleBeamMaps}, or similarly \textit{jet transport} \cite{Yuan2024JetTransport}, provides a computational representation of the local Taylor expansion of a nonlinear map. Let $\bm{g}:\mathbb{R}^N\rightarrow\mathbb{R}^m$ be sufficiently smooth near a reference state $\bm{X}_r$, and let $\delta\bm{X}=\bm{X}-\bm{X}_r$. Using a multi-index notation $\bm{\alpha}=(\alpha_1,\ldots,\alpha_N)$, define
\begin{equation}
\label{eq:multi_index_definitions}
|\bm{\alpha}|=\sum_{i=1}^N\alpha_i,\qquad
\bm{\alpha}! = \prod_{i=1}^N \alpha_i!,\qquad
\delta\bm{X}^{\bm{\alpha}}=\prod_{i=1}^N \delta X_i^{\alpha_i}.
\end{equation}
The order-$j$ Taylor map can then be identified from
\begin{equation}
\label{eq:da_multi_index_expansion}
\bm{g}(\bm{X}_r+\delta\bm{X})
=
\bm{g}(\bm{X}_r)
+
\sum_{1\leq |\bm{\alpha}|\leq j}
\frac{1}{\bm{\alpha}!}
D^{\bm{\alpha}}\bm{g}(\bm{X}_r)
\delta\bm{X}^{\bm{\alpha}}
+
\text{H.O.T.}
\end{equation}
Under usual smoothness assumptions, mixed partial derivatives are invariant to permutation of the differentiation order. Consequently, the number of unique derivative coefficients at a given order is exactly the number of unique monomials at that order. For an expansion through order $j$ in $N$ variables, the number of monomials is therefore
\begin{equation}
\label{eq:full_da_monomial_count}
K(N,j)=
\binom{N+j}{N}-1.
\end{equation}
This combinatorial growth motivates directional DA: rather than constructing the full Taylor map in all variables, one may choose DA independent variables adapted to physically important directions, retaining nonlinear dependence only in selected coordinates.

\subsection{Directional Differential Algebra}

Directional differential algebra (DDA) is our term for a targeted implementation of standard DA in which the independent variables are chosen to align with the directions of interest in the problem. Since DA maps may be constructed with respect to any convenient coordinates, one may replace the full perturbation $\delta\bm{X}\in\mathbb{R}^N$ with a directional decomposition
\begin{equation}
\label{eq:dda_directional_decomposition}
\delta\bm{X}=\hat{\bm{\gamma}}^*\chi+L\bm{\epsilon},
\end{equation}
where $\hat{\bm{\gamma}}^*$ is a prioritized unit direction, $\chi$ is the corresponding scalar deviation measuring along $\hat{\bm{\gamma}}^{*}$, and $L\bm{\epsilon}$ represents departures along the remaining orthogonal directions $L[:,i]\cdot\hat{\bm{\gamma}}^{*}=0$ with associated coordinates $\bm{\epsilon}\in\mathbb{R}^{N-1}$. Existing DA packages (e.g., DACE \cite{RasottoDACE}, used in this work) easily accommodate such an approach. A directional order-$j$ approximation of a nonlinear map $\bm{g}$ at $\bm{X} = \bm{X}_{r} + \delta\bm{X}$ is then written as
\begin{equation}
\label{eq:dda_directional_expansion}
\begin{split}
\bm{g}(\bm{X})
& \approx
\bm{g}(\bm{X}_r)
+
\sum_{q=1}^{j}
\frac{1}{q!}
D^q\bm{g}(\bm{X}_r)
\left[
\hat{\bm{\gamma}}^*,\ldots,\hat{\bm{\gamma}}^*
\right]\chi^q \\
& +
\frac{\partial \bm{g}}{\partial \bm{X}}(\bm{X}_r)L\bm{\epsilon}.
\end{split}
\end{equation}
Thus, DDA retains the pure nonlinear dependence along $\hat{\bm{\gamma}}^*$ while keeping only a first-order correction in the transverse coordinates. This reduces the number of retained non-constant terms from that of Eq.~\eqref{eq:full_da_monomial_count} to $N+j-1$ terms, preserving the dominant nonlinear structure in a cylindrical neighborhood aligned with $\hat{\bm{\gamma}}^*$. It is also possible to retain more than one direction, but the mixing terms should be computed in that case.
\subsection{Uses of DA in UQ Methods}
In lieu of integrating the nonlinear dynamics for every sample or sigma point, we can apply DA or DDA to provide a nonlinear Taylor series expansion of the flow across the domain of interest. This step can be done offline or in advance of the UQ step, which significantly reduces real-time computational effort whenever the UQ method is to be employed. The trade-off is that DA requires that the dynamic model satisfy certain smoothness properties, and that any relevant code admits the use of overloaded operators to accommodate the differential terms \cite{Berz_ParticleBeamMaps, Yuan2024JetTransport}.

In lieu of a roundabout application of DA to the sigma point-based methods for estimating statistical moments, we note that it is also possible to recover the statistical moments directly from the DA map. This was explored in Reference~\cite{valli_nonlinear_2013}. They note an explosion in the number of monomial terms in the computation of full higher-order tensors. For instance, the fourth-order kurtosis tensor, when computed with a third-order expansion, involves the expectations of monomials up to order 12. This can quickly become computationally quite burdensome with higher-order maps. References \cite{acciarini_nonlinear_2025} and~\cite{LefebvrePCE} provide more thorough discussions of such matters, along with techniques for exploiting sparsity and redundancy of the expansion computations to reduce the burden. We save exploration of these techniques for upcoming work, but note that many UQ applications require computation of only some components or low-dimensional projections of the entire higher-order moment tensor. Thus one can avoid computing many components for numerical efficiency.

\subsection{Non-Gaussian Confidence Boundary Methods}
\subsubsection{Analytic Banana Contour}
To demonstrate the use of higher-moment computation, we employ the CUT4-computed
third- and fourth-order central moments to construct an analytic approximation of a
non-Gaussian confidence boundary. The method follows Ref.~\cite{BurnettBooneCDC26},
where a non-Gaussian confidence contour is treated as a perturbation of the Gaussian
confidence ellipse. Here we repeat only the resulting construction. We recommend a careful read of Ref.~\cite{BurnettBooneCDC26} to the interested reader.

Consider a $2D$ slice $\bm{r}_q\in\mathbb{R}^2$ of the propagated random
state $\bm{X}_f$, where $q$ denotes the slice state indices, with mean $\bm{\mu}_q$, covariance $\Sigma_q$, and higher-order central moment tensors $\mathcal{M}^{(3)}_q$ and $\mathcal{M}^{(4)}_q$.
Let
\begin{equation}
\label{eq:ngcb_covariance_eigendecomp}
\Sigma_q
=
R_q\Lambda_qR_q^\top,
\qquad
\Lambda_q
=
\mathrm{diag}(\lambda_{1,q},\lambda_{2,q}),
\end{equation}
where $\lambda_{1,q}\geq \lambda_{2,q}>0$. For a confidence scaling $k$, the Gaussian
confidence contour is
\begin{equation}
\label{eq:ngcb_gaussian_contour}
\bm{r}_{q,G}(t)
=
\bm{\mu}_q
+
R_q
\begin{pmatrix}
k\sqrt{\lambda_{1,q}}\cos t\\
k\sqrt{\lambda_{2,q}}\sin t
\end{pmatrix},
\qquad
t\in[0,2\pi).
\end{equation}
This is fully determined by the mean and covariance. To include moderate
non-Gaussian deformation, we first introduce the whitened principal-axis coordinates
\begin{equation}
\label{eq:ngcb_whitened_coordinates}
\begin{pmatrix}
\hat{u}\\
\hat{v}
\end{pmatrix}
=
W_q(\bm{r}_q-\bm{\mu}_q),
\qquad
W_q
=
\Lambda_q^{-1/2}R_q^\top .
\end{equation}
Let $\bm{a}_q^\top$ and $\bm{b}_q^\top$ denote the first and second rows of
$W_q$. The quantities used
in the boundary correction are
\begin{equation}
\label{eq:ngcb_projected_moments}
\begin{split}
m_{uuu,q}
&=
\sum_{i,j,k}
a_{q,i}a_{q,j}a_{q,k}
\mathcal{M}^{(3)}_{q,ijk},\\
m_{uuv,q}
&=
\sum_{i,j,k}
a_{q,i}a_{q,j}b_{q,k}
\mathcal{M}^{(3)}_{q,ijk},\\
m_{uuuu,q}
&=
\sum_{i,j,k,l}
a_{q,i}a_{q,j}a_{q,k}a_{q,l}
\mathcal{M}^{(4)}_{q,ijkl}.
\end{split}
\end{equation}
The first correction captures the transverse bend of a banana-shaped distribution. It is
written as a quadratic correction in the short-axis coordinate, with coefficient
\begin{equation}
\label{eq:ngcb_bend_coefficient}
\alpha_q
=
\frac{m_{uuv,q}}{m_{uuuu,q}-1}.
\end{equation}
The second correction captures long-axis asymmetry using the first-order Cornish-Fisher
correction,
\begin{equation}
\label{eq:ngcb_cf_coefficient}
c_q(k)
=
\frac{k^2-1}{6}\,m_{uuu,q}.
\end{equation}
Combining these two corrections gives the perturbed principal-axis values of the confidence contour
\begin{equation}
\label{eq:ngcb_corrected_uv}
\begin{split}
u_q(t)
&=
k\sqrt{\lambda_{1,q}}\cos t
+
c_q(k)\sqrt{\lambda_{1,q}}\cos^2 t,\\
v_q(t)
&=
k\sqrt{\lambda_{2,q}}\sin t
+
\alpha_q\sqrt{\lambda_{2,q}}
\left(k^2\cos^2 t-1\right).
\end{split}
\end{equation}
The corresponding non-Gaussian confidence contour in the original slice coordinates is
then
\begin{equation}
\label{eq:ngcb_corrected_contour}
\bm{r}_{q,\mathrm{NG}}(t)
=
\bm{\mu}_q
+
R_q
\begin{pmatrix}
u_q(t)\\
v_q(t)
\end{pmatrix},
\qquad
t\in[0,2\pi).
\end{equation}

Equation~\eqref{eq:ngcb_projected_moments} need not be evaluated by first forming the
full tensors $\mathcal{M}^{(3)}_q$ and $\mathcal{M}^{(4)}_q$. If a polynomial map
$\bm{T}_q(\bm{\zeta})$ is available for the slice, where $\bm{\zeta}$ denotes the
uncertain input variables used by the chosen UQ representation, define
\begin{equation}
\label{eq:projected_scalar_maps}
U_q=\bm{a}_q^\top(\bm{T}_q-\bm{\mu}_q),
\qquad
V_q=\bm{b}_q^\top(\bm{T}_q-\bm{\mu}_q).
\end{equation}
Then the required projected moments are e.g.
\begin{equation}
\label{eq:projected_scalar_moments}
m_{uuu,q}=\mathbb{E}[U_q^3],
\qquad
m_{uuv,q}=\mathbb{E}[U_q^2V_q].
\end{equation}
Thus the map is projected before explicit moment computations, avoiding construction of moment-tensor entries not explicitly used by the boundary model. 

The first four moments do not uniquely determine an
arbitrary non-Gaussian confidence boundary; rather, Eq.~\eqref{eq:ngcb_corrected_contour}
provides a minimum-complexity approximation. Nonetheless this approach has performed well in diverse numerical studies, providing a fully analytic confidence boundary with runtime well within the same order of magnitude as LinCov in our experiments.

\section{Numerical Studies}
\subsection{UQ Method Trade Study}
For our first numerical test, we compare a collection of standard sample- and sigma-point-based uncertainty propagation methods applied both directly with the nonlinear dynamics and with the pre-computed DA/DDA representations of the flow map. The nominal trajectory is taken from a JPL SSD southern \(L_2\) halo orbit in the Earth--Moon circular restricted three-body problem (see e.g. \cite{Koon:2006rf}), with Jacobi constant \(C_{J}=3.0612627924\) and period \(T=3.136654204\) nondimensional time units (TUs), or approximately \(13.62\) days. 

\begin{figure}[h!]
\centering
\includegraphics[width=0.95\columnwidth]{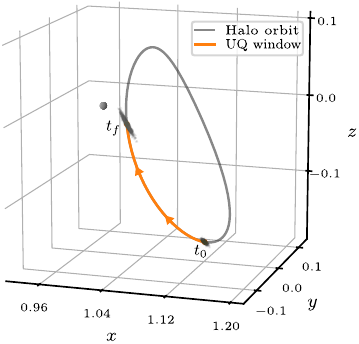}
\caption{\itshape Earth--Moon \(L_2\) halo orbit, UQ arc, and dispersed samples}
\label{fig:haloUQ}
\end{figure}

The UQ study is initialized 0.25 TU from apolune, which is then numerically integrated for 0.9 TU, which provides a moderately nonlinear, three-dimensional trajectory for assessing the accuracy and runtime tradeoffs of the competing UQ representations. 
The initial uncertainty is imposed on the deviated initial state
\(\delta\bm{X}_0=\bm{X}_0-\bm{X}_r\). The deviation mean and covariance are (in non-dimensional units)
\begin{equation}
    \bar{\delta\bm{X}}_0 =
    \begin{bmatrix}
    0 & 10^{-4} & 0 & 0 & 10^{-4} & 0
    \end{bmatrix}^{\top},
    \label{eq:cr3bp_initial_deviation_mean}
\end{equation}
\begin{equation}
    P_0 = 10^{-6} I_6 + 10^{-5}\hat{\bm{\gamma}}^{*}
    \left(\hat{\bm{\gamma}}^{*}\right)^{\top},
    \label{eq:cr3bp_initial_covariance}
\end{equation}
where \(\hat{\bm{\gamma}}^{*}\) is the preferential direction used by the DDA
construction. This is selected from the finite-time linearized dynamics
over the UQ interval, with local time \(t_0=0\) and
\(t_f=\Delta t_{\rm UQ}=0.9\) TU. After propagating the STM
\(\Phi(t_f,t_0)\) along the nominal trajectory, we form
\begin{equation}
    G(t_f,t_0)
    =
    \Phi(t_f,t_0)^{\top}\Phi(t_f,t_0),
    \label{eq:cauchy_green_direction}
\end{equation}
and choose \(\hat{\bm{\gamma}}^{*}\) as the normalized eigenvector associated
with the largest eigenvalue of \(G\). Thus,
\(\hat{\bm{\gamma}}^{*}\) is the initial perturbation direction that undergoes
the largest linearized stretching over the UQ arc. 
A plot of the Monte Carlo samples from the initial time and final time is given in Figure \ref{fig:haloUQ}, in CR3BP normalized length units.

Table \ref{table:ex1_runtime_compact} summarizes runtime performance for all methods. This is done in a Python test script on a 2024 MacBook Pro with Apple M4 Max chip, using \texttt{daceypy}, a Python wrapper of DACE \cite{RasottoDACE}. Table~\ref{table:ex1_runtime_compact} shows that direct MC is by far the most expensive option, requiring \(12.786~{\rm s}\) for \(10^4\) samples in this simple and relatively short-duration scenario. Deterministic sigma-point methods are much cheaper, with UT and CUT4 running notably faster than the GMM and PCE. The DA and DDA maps introduce a one-time construction cost, \(0.292~{\rm s}\) and \(0.139~{\rm s}\), but greatly reduce repeated evaluation costs. 

For all pre-computed map evaluations, DDA is consistently faster than full DA. However, there is an intuitive tradeoff to be aware of for the DDA approach: while it retains the dominant higher-order variation along a preferential direction, it neglects all higher-order ``off-direction" terms, which carries some reduction in accuracy, especially by some metrics. For instance, Table~\ref{table:ex1_errors_compact} illustrates this effect with UT. The covariance error in the table is reported as a relative Frobenius error,
\begin{equation}
    \mathrm{err}_{P_f}
    =
    \frac{
    \left\|P_f^{\mathrm{method}} - P_f^{\mathrm{UT}}\right\|_F
    }{
    \left\|P_f^{\mathrm{UT}}\right\|_F
    }.
    \label{eq:covariance_error_metric}
\end{equation}
This metric penalizes discrepancies in both marginal variances and mixed covariance terms. By this metric, comparing the DA and DDA results, we see higher relative covariance error compared with the mean error, consistent with the fact that off-direction and off-diagonal covariance accuracy is more strongly affected by the directional truncation. It still improves on the LinCov result, however. 
Thus, DDA trades some higher-order moment fidelity for a substantial reduction in map construction and evaluation cost.

Regarding relative accuracy of the various standard methods, we note that only MC, PCE, CUT4, and GMM can provide information about higher-order moments. MC methods have extremely sluggish convergence in moment estimation with respect to the number of random samples. The CUT4 implementation and PCE give similar predictions for higher-order moments in this study, but CUT4 has the advantage of being deterministic, in contrast to the non-intrusive PCE implementation used here. Beyond the preceding comments, we omit a lengthy numerical comparison of UQ method relative accuracies due to length constraints.

For the GMM, an initial Gaussian is split deterministically into a small mixture by recursively dividing each component along its dominant
covariance eigenvector. Each split produces two equal-weight child components with shifted means and reduced covariance, preserving the parent mean and covariance. In this particular test, we have 16 components, each of which is propagated using the UT. No optimized mixture-weight or anchor-selection procedure is included -- the GMM is tuned manually, hence the ``N/A" entry for the GMM weight computation in Table \ref{table:ex1_runtime_compact}. We include this result only for completeness.

\begin{table}[h!]
\centering
\caption{Runtime comparison for selected UQ methods.}
\label{table:ex1_runtime_compact}
\setlength{\tabcolsep}{3pt}
\renewcommand{\arraystretch}{0.92}
\begin{tabularx}{\columnwidth}{@{}lXr@{}}
\toprule
Method & Operation & $dt$ [s] \\
\midrule
\multicolumn{3}{@{}l}{\emph{Direct propagation}} \\
MC     & $10^4$ samples      & 12.786 \\
LinCov & STM/covariance      & 0.005 \\
UT     & sigma points        & 0.017 \\
CUT4   & sigma points        & 0.097 \\
PCE    & total runtime       & 2.928 \\
GMM    & mixture             & 0.269$^\dagger$ \\
\midrule
\multicolumn{3}{@{}l}{\emph{Representation construction}} \\
DA     & full third-order map        & 0.292 \\
DDA    & directional third-order map & 0.139 \\
PCE    & basis generation        & 0.058 \\
GMM    & initial weights             & N/A$^{\dagger}$ \\
\midrule
\multicolumn{3}{@{}l}{\emph{Precomputed representation evaluation}} \\
DA+MC     & $10^4$ map calls       & 0.937 \\
DDA+MC    & $10^4$ map calls       & 0.061 \\
DA+UT     & UT map calls           & $1.28{\times}10^{-3}$ \\
DDA+UT    & UT map calls           & $1.35{\times}10^{-4}$ \\
DA+CUT4   & CUT4 map calls         & $7.91{\times}10^{-3}$ \\
DDA+CUT4  & CUT4 map calls         & $1.28{\times}10^{-3}$ \\
DA+PCE    & PCE fit from map calls & 0.234 \\
DDA+PCE   & PCE fit from map calls & 0.098 \\
DA+GMM    & mixture anchors        & 0.020 \\
\bottomrule
\end{tabularx}
\end{table}
\begin{table}[h!]
\centering
\caption{Errors vs. nonlinear Unscented Transform.}
\label{table:ex1_errors_compact}
\setlength{\tabcolsep}{4pt}
\renewcommand{\arraystretch}{0.95}
\begin{tabularx}{\columnwidth}{@{}Xcc@{}}
\toprule
Method & $\lVert \mathrm{err}_{\mu} \rVert$ & $\mathrm{err}_{P_f}$ \\
\midrule
Full DA+UT & $3.36{\times}10^{-8}$ & $2.24{\times}10^{-7}$ \\
DDA+UT     & $3.32{\times}10^{-5}$ & $1.498{\times}10^{-3}$ \\
LinCov     & $1.820{\times}10^{-3}$ & $1.035{\times}10^{-2}$ \\
\bottomrule
\end{tabularx}
\end{table}

\subsection{Non-Gaussian Confidence Analysis}
As another test case of the methods discussed in this work, consider an Earth-return aerocapture problem, whereby a hyperbolic trajectory is captured by a single carefully managed atmospheric braking event. In the case of Earth aerocapture, such a scenario has been proposed in the past for sample return, or as part of the Earthbound leg of a sustainable cislunar architecture. 

\begin{figure}[h!]
\centering
\includegraphics[width=0.9\columnwidth]{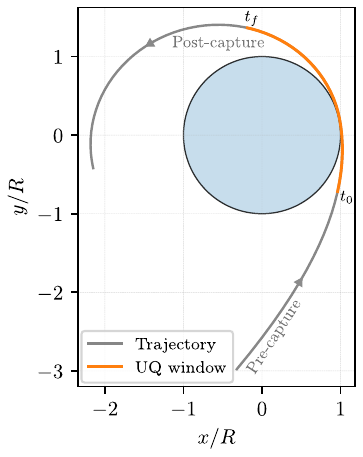}
\caption{\itshape Aerocapture nominal trajectory}
\label{fig:aerocap}
\end{figure}

Figure~\ref{fig:aerocap} shows the nominal aerocapture trajectory. The vehicle arrives on a hyperbolic Earth approach with \(v_\infty=2.5~{\rm km/s}\) and \(e_{\rm in}=1.1009\), enters the atmospheric interface at \(h_E=100~{\rm km}\), and exits after \(239.7~{\rm s}\) of atmospheric flight onto a bound orbit with \(e_f=0.3698\) and apoapsis altitude \(r_a-R_E=7.60\times10^3~{\rm km}\). The atmosphere is modeled as \(\rho(h)=\rho_E\exp[-(h-h_E)/H]\), with
\(\rho_E=5.0\times10^{-7}~{\rm kg/m^3}\), scale height \(H=7.2~{\rm km}\), and ballistic coefficient \(\beta=500~{\rm kg/m^2}\).

\begin{figure}[h!]
\centering
\includegraphics[width=0.95\columnwidth]{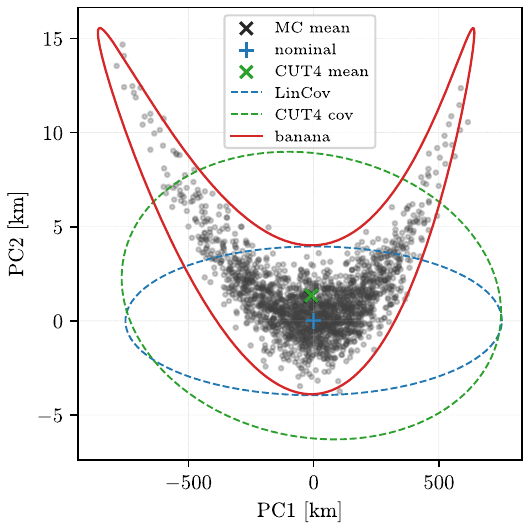}
\caption{\itshape Post-aerocapture 3$\sigma$ confidence predictions}
\label{fig:aeroUQ}
\end{figure}

Consider the dispersion Case 5 in Table~\ref{tab:aerocap_uq_cases}. Figure~\ref{fig:aeroUQ} shows the resulting predicted 3$\sigma$ confidence boundary in position, in a principal coordinate frame aligned with the LinCov solution, along with a small-batch Monte Carlo sampling of 2000 points. The Monte Carlo samples illustrate a non-Gaussian final distribution, greatly distorted from the initially tight Gaussian distribution. The axes are rescaled to show the non-Gaussian bending effect, as aerocapture problems strongly shear the distribution along a maximum stretching direction. 

In Figure~\ref{fig:aeroUQ}, whereas the LinCov $3\sigma$ boundary encompasses 88.6\% of the sample points, the banana contour encompasses 99.55\%. This is
slightly conservative relative to the expected two-dimensional $3\sigma$ coverage of approximately 98.9\%, while providing a much more accurate spatial characterization of the non-Gaussian distribution.
We note also that the entire banana contour UQ characterization required 0.227s in our Python test script. The LinCov solution took 0.048s, so the banana contour characterization took only 4.7 times as long as LinCov, without any use of DA. The 2000 Monte Carlo samples took 18.6s to propagate.

Tables~\ref{tab:aerocap_uq_cases} and~\ref{tab:aerocap_uq_runtime} summarize the results of other aerocapture UQ test cases, with comparisons to 4000 sample Monte Carlo, illustrating robustness of the banana contour approach. All cases have an entry flight path angle (EFPA) of -4.85$^\circ$, except case 7, which applies a slightly steeper EFPA of -4.87$^\circ$. The parameters $t_{\text{pre}}$ and $\Delta t_{\text{UQ}}$ report the initialization time in seconds before atmospheric entry and the time duration of the UQ study, respectively. Initial uncertainties are defined in initial radial and transverse components. The banana method outperforms LinCov in all cases and CUT4 covariance ellipse in 5 of 7, with greatly reduced over-conservatism.
\begin{table}[h!]
\centering
\setlength{\tabcolsep}{2.5pt}
\caption{Aerocapture UQ cases}
\label{tab:aerocap_uq_cases}
\begin{tabular}{@{}lrrrr@{}}
Case No. & \(t_{\rm pre}\) & \(\Delta t_{\rm UQ}\) &
\(\sigma_{R},\sigma_{T}\) & \(\sigma_{V_R}, \sigma_{V_T}\) \\
 & [s] & [s] & [m] & [m/s] \\
\hline
1 & 208 & 649  & 30, \ 800  & 0.030, \ 0.6 \\
2 & 418 & 1868 & 30, \ 800  & 0.030, \ 0.6 \\
3 & 208 & 730  & 15, \ 400  & 0.015, \ 0.3 \\
4 & 208 & 649  & 15, \ 400  & 0.015, \ 0.3 \\
5 & 418 & 1868 & 15, \ 400  & 0.015, \ 0.3 \\
6 & 208 & 649  & 19.5, \ 520 & 0.0195, \ 0.39 \\
7 & 208 & 657  & 15, \ 400  & 0.015, \ 0.3 \\
\hline
\end{tabular}
\end{table}
\begin{table}[h!]
\centering
\setlength{\tabcolsep}{2.6pt}
\caption{Runtime and Monte Carlo coverage by method}
\label{tab:aerocap_uq_runtime}
\begin{tabular}{@{}lrrrr@{}}
Case No. & \(t_{\rm MC}\) & \(t_{\rm LC}\) & \(t_{\rm banana}\) & Monte Carlo coverage \\
 & [s] & [s] & [s] & [\% LC, CUT4, ban.] \\
\hline
1 & 13.12 & 0.020 & 0.082 & 76.4, \ 97.6, \ 95.8 \\
2 & 36.46 & 0.049 & 0.229 & 75.7, \ 97.7, \ 97.1 \\
3 & 14.36 & 0.022 & 0.090 & 87.4, \ 98.0, \ 99.3 \\
4 & 12.48 & 0.019 & 0.077 & 88.4, \ 98.0, \ 99.4 \\
5 & 36.39 & 0.048 & 0.229 & 88.8, \ 98.0, \ 99.6 \\
6 & 12.96 & 0.020 & 0.081 & 84.2, \ 97.8, \ 98.7 \\
7 & 13.24 & 0.020 & 0.082 & 84.6, \ 97.8, \ 99.2 \\
\hline
\end{tabular}
\end{table}

\section{Discussion}
This work compares nonlinear UQ strategies for spaceflight problems, emphasizing
the runtime benefits of precomputed DA/DDA flow maps. The CR3BP study shows that
such maps make repeated MC, sigma-point, PCE, and GMM evaluations much cheaper
than repeated nonlinear propagation. DDA further reduces this cost by retaining
nonlinear dependence mainly along a preferential stretching direction, at the
expected expense of off-direction fidelity.

The aerocapture study illustrates the potential importance of higher-order moments in highly nonlinear regimes: a LinCov approach can misrepresent the propagated
distribution, while the CUT4-based banana approximation uses selected third- and fourth-order moments to recover a much more representative confidence boundary at
modest additional cost. These results also suggest a path toward stochastic extensions of recently proposed fast localized guidance methods based on DA flow maps
\cite{BurnTopp2025, RegantiniJGCD}. Specifically, map-based UQ could supply the confidence boundary information needed for chance-constrained variants.

\section{Acknowledgements}
The authors gratefully acknowledge helpful discussions with Mr. Niccolò Michelotti at Politecnico di Milano.
\section{References}

%
\bibliographystyle{issfd2026}
\bibliography{references}

\end{document}